\documentclass{article}

\usepackage{style} 
\usepackage[utf8]{inputenc} 
\usepackage[T1]{fontenc}    
\usepackage{url}            
\usepackage{booktabs}       
\usepackage{amsfonts}       
\usepackage{nicefrac}       
\usepackage{microtype}      
\usepackage{graphicx}
\usepackage{booktabs}       
\usepackage{amsfonts}       
\usepackage{nicefrac}       
\usepackage{microtype}      
\usepackage{multirow}
\usepackage{multicol}
\usepackage{wrapfig}
\usepackage{graphicx}
\usepackage{booktabs} 
\usepackage{makecell}
\usepackage{amsmath}
\usepackage{amssymb}
\usepackage{comment}
\usepackage{color}
\usepackage{graphicx}
\usepackage{amsmath}
\usepackage{amssymb}
\usepackage{booktabs}
\usepackage{algorithm}
\usepackage{algpseudocode}
\usepackage{wrapfig}
\usepackage{lipsum}
\usepackage{wrapfig}
\usepackage{mathtools}
\usepackage{bm}
\usepackage{enumitem}
\usepackage{bbding}
\usepackage{bbm}
\usepackage{caption}
\usepackage{varwidth}
\usepackage{fontawesome}
\usepackage{bbding}
\usepackage[export]{adjustbox}
\usepackage{footnote}
\usepackage{import}

\DeclareMathOperator*{\argmin}{arg\,min}

\DeclareMathAlphabet{\pazocal}{OMS}{zplm}{m}{n}

\DeclareMathAlphabet\mathbfcal{OMS}{cmsy}{b}{n}
\usepackage[table,dvipsnames]{xcolor}
\definecolor{topcolor}{rgb}{1,0.8,0.8}
\definecolor{secondcolor}{rgb}{1,0.87,0.7}
\definecolor{thirdcolor}{rgb}{1,1,0.8}

\hypersetup{
linkcolor=BrickRed
,citecolor=Green
,filecolor=Mulberry
,urlcolor=NavyBlue
,menucolor=BrickRed
,runcolor=Mulberry
,linkbordercolor=BrickRed
,citebordercolor=Green
,filebordercolor=Mulberry
,urlbordercolor=NavyBlue
,menubordercolor=BrickRed
,runbordercolor=Mulberry
}

\title{SPEAR\includegraphics[scale=0.32]{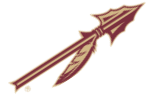}: Receiver-to-Receiver Acoustic\\ Neural Warping Field}

\author{
 Yuhang He \footnotemark[2]\ \ \ \    Shitong Xu \footnotemark[2]\\
 Jia-Xing Zhong\ \ \ \ \  Sangyun Shin\ \ \ \    Niki Trigoni\ \ \ \    Andrew Markham \\
  Department of Computer Science\\
  University of Oxford, Oxford, United Kingdom.\\
  \faGithub\  \url{https://github.com/xu-shitong/SPEAR}\\
   \faEnvelopeO\ \texttt{yuhang.he@cs.ox.ac.uk, shitong.xu@cs.ox.ac.uk}\\
}
\begin{document}
\maketitle

\footnotetext[1]{ Yuhang He and Shitong Xu are of equal contributions.}

\begin{abstract}
We present \emph{SPEAR}, a continuous receiver-to-receiver acoustic neural warping field for spatial acoustic effects prediction in an acoustic 3D space with a single stationary audio source. Unlike traditional source-to-receiver modelling methods that require prior space acoustic properties knowledge to rigorously model audio propagation from source to receiver, we propose to predict by warping the spatial acoustic effects from one reference receiver position to another target receiver position, so that the warped audio essentially accommodates all spatial acoustic effects belonging to the target position. \emph{SPEAR} can be trained in a data much more readily accessible manner, in which we simply ask two robots to independently record spatial audio at different positions. We further theoretically prove the universal existence of the warping field if and only if one audio source presents. Three physical principles are incorporated to guide \emph{SPEAR} network design, leading to the learned warping field physically meaningful. We demonstrate \emph{SPEAR} superiority on both synthetic, photo-realistic and real-world dataset, showing the huge potential of \emph{SPEAR} to various down-stream robotic tasks.
\end{abstract}
\vspace{-2mm}
\vspace{-2mm}
\section{Introduction}
\vspace{-2mm}

In an enclosed acoustic 3D space where a stationary sound source keeps emitting spatial audio, the primary objective is to precisely delineate the spatial acoustic effects for any given receiver position. These spatial acoustic effects typically encompass reverberation, loudness variation and resonance. Achieving high-fidelity and authentic spatial acoustic effect modelling is pivotal for delivering a truly immersive 3D acoustic experience that seamlessly integrates with the 3D room scene. Consequently, such modelling techniques have a wide range of applications in auditory AR/VR techniques~\cite{3Dimmersive,immersive_spatialaudio,game_audio}, audio-inclusive robot tasks~\cite{a_slam} and reconstruction endeavors~\cite{acousNet,boombox}. 

To model the spatial acoustic effects, most prior methods~\cite{bilbao2017wave-based,georoom_overview,ISM,raytracing,pietrzyk1998computer,kleiner1993auralization-an,finitediff,exp_radiosity,NAF} follow the source-to-receiver pipeline to explicitly model sound propagation process from source to receiver, where the overall behavioural change along the propagation path is usually described by room impulse response~(RIR). As the spatial audio propagates in a complex way encompassing diffraction, reflection and absorption, the resulting RIR is highly non-smooth and lengthy in data points. Classic methods, either wave-based~\cite{bilbao2017wave-based,pietrzyk1998computer,kleiner1993auralization-an,finitediff} and geometry-based modelling~\cite{georoom_overview}, require massive prior knowledge of the 3D space's acoustic properties such as source position, space geometric layout and constructional material to precisely simulate the propagation process for a given source-receiver pair. However, accessing such prior knowledge poses a formidable challenge in reality and the whole computation is inextricably inefficient. Some recent work~(\textit{e.g.}, NAF~\cite{NAF}) aim to alleviate this computational burden by learning a continuous acoustic neural field. Nonetheless, training such a continuous field necessitates vast RIR data which is exceedingly difficult to collect in real scenarios.

In this work, we instead propose to predict spatial acoustic effects from receiver-to-receiver perspective. Our framework, termed \emph{SPEAR}, relies on neither RIR data nor prior space acoustic properties that are difficult to obtain and required by existing source-to-receiver based methods~\cite{NAF,georoom_overview,bilbao2017wave-based}, but instead simply require much more readily accessible data -- the receiver recorded audio at discrete positions. Our observation is that directly carrying a receiver~(can ask robot or human to hold the receiver) to record audio at different positions is much readily executable than measuring RIR data and space acoustic properties. Since receiver recorded audio naturally encodes the spatial acoustic effects at its position, analyzing the received audio can help to acoustically characterize the 3D space, and further predict the spatial acoustic effects for any given novel receiver position.

\begin{figure*}[t]
    \centering
    \includegraphics[width=0.98\linewidth]{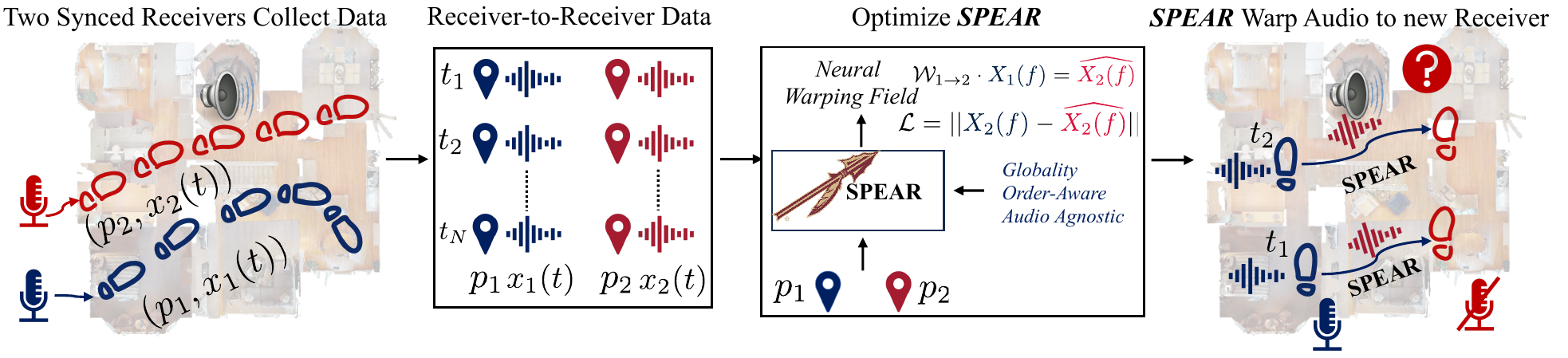}
    \caption{\textbf{SPEAR Motivation}: A stationary audio source is emitting audio in 3D space. Requiring neither source position nor 3D space acoustic properties, \emph{SPEAR} simply requires two microphones to actively record the spatial audio independently at discrete positions. During training, \emph{SPEAR} takes as input a pair of receiver positions and outputs a warping field potentially warping the recorded audio on reference position to target position. Minimizing the discrepancy between the warped audio and recorded audio enforces \emph{SPEAR} to acoustically characterise the 3D space from receiver-to-receiver perspective. The learned \emph{SPEAR} is capable of predicting spatial acoustic effects at arbitrary positions.}
    \label{fig:spear_mot}
    \vspace{-3mm}
\end{figure*}

As is shown in Fig.~\ref{fig:spear_mot}, to obtain the training data, we simply require receivers to record the audio in the 3D space independently. At each discrete position pair, the two receivers are temporally synchronized to record the same audio content and their respective positions are recorded as well. \emph{SPEAR} then learns a continuous receiver-to-receiver acoustic neural warping field that takes as input two receivers' position and outputs a neural warping field warping the spatial acoustic effects from one reference position to the other target position. With the learned \emph{SPEAR}, we can warp the audio recorded at one arbitrary reference position to another arbitrary target position so that the warped audio fully accommodates the spatial acoustic effects at the target position. \emph{SPEAR} has huge potential in various robotic tasks, such as audio-involved robot relocalization and manipulation.

We further theoretically prove the universal existence of the receiver-to-receiver warping field if and only if one stationary audio source presents in the 3D space, then we introduce three main physical principles underpinned by linear time-invariant~(LTI) 3D acoustic space that guide \emph{SPEAR} neural network design: \emph{Globality}, \emph{Order Awareness} and \emph{Audio-Content Agnostic}. We adopt the Transformer~\cite{att_all_need} architecture to predict the warping field in frequency domain, where the lengthy warping field is divided into small and non-overlapping patches. Each token is responsible for predicting a patch. We run experiments on both synthetic, photo-realistic and real-world datasets to show the superity of \emph{SPEAR}. In summary, we make three main contributions:

\vspace{-3mm}
\begin{enumerate}[leftmargin=*]
    \item We propose \emph{SPEAR}, a novel receiver-to-receiver spatial acoustic effects prediction framework. Unlike existing source-to-receiver modelling methods requiring extensive prior space acoustic properties knowledge, \emph{SPEAR} can be efficiently trained in a data more readily accessible manner.
    \item We theoretically prove the universal existence of receiver-to-receiver neural warping field if and only if one stationary audio source presents in the 3D space. \emph{SPEAR} network design is based on three acoustic physical principles, so that the whole neural network is physically meaningful.
    \item We demonstrate \emph{SPEAR} superiority on both synthetic data, photo-realistic and real-world dataset.
\end{enumerate}
\vspace{-2mm}
\section{Related Work}
\vspace{-2mm}

\textbf{Spatial Acoustic Effects Modelling.} Classic methods tend to numerically compute room impulse response~(RIR). They can be divided into two main categories: wave-based~\cite{bilbao2017wave-based,pietrzyk1998computer,kleiner1993auralization-an,finitediff} and geometry-based~(aka geometrical acoustics)~\cite{georoom_overview,raytracing,ISM,beamtracing,exp_radiosity,inv_radiosity}. They require extensive prior knowledge of the 3D space acoustic properties such as the audio source position~\cite{sounddet,soundsynp}, room geometry and furniture arrangement to derive RIR. Moreover, they are computationally expensive and the whole computation needs to be resumed once either the source or receiver position gets changed. Our proposed \emph{SPEAR} circumvents the dependency on extensive prior knowledge and learns the spatial acoustic effects in a data more readily accessible manner. The advent of deep neural networks has inspired recent work~\cite{fast_rir,diff_acousgen,IR-GAN,TS-RIR,SDNet,NAF,richard2022deepimpulse,majumder2022fewshot,improved_RIR,avrir} to focus on learning RIR with deep neural networks. While showing promising performance, they still require massive RIR data or even crossdomain visual data~\cite{avrir} to train their model, which in reality are difficult to collect. While all of those methods fall into source-to-receiver estimation, \emph{SPEAR} infers spatial acoustic effects from a receiver-to-receiver perspective which naturally brings several advantages over existing methods.

\textbf{Neural Implicit Representation.} Implicit representation learning has received lots of attention in recent years, especially in computer vision community~\cite{mildenhall2020nerf,hedman2021snerg,HNeRF,ANeRF,pixelnerf}. They model static or dynamic visual scenes by optimizing an implicit neural radiance field in order to render photo-realistic novel views. Some recent work~(\textit{e.g.}, NAF~\cite{NAF}, FewShotRIR~\cite{majumder2022fewshot}) propose to learn implicit neural acoustic fields from source-receiver pairs or audio-visual cues. \emph{SPEAR} also learns a spatial continuous neural implicit representation for spatial acoustic effects prediction.

\textbf{Audio Synthesis}. Estimating the audio for at a novel position partially relates to audio synthesis~\cite{wavenet,donahue2019wavegan,zuiderveld2021towards,richard2021binaural,GANSynth,spatialsreprod,diffimpact,waveglow}. WaveNet~\cite{wavenet} learns to predict future sound waveform based on previously heard sound waveform. WaveGAN~\cite{donahue2019wavegan} and GANSynth~\cite{GANSynth} adopt generative adversarial network~(GAN~\cite{GAN}) to learn to generate audio. Our framework \emph{SPEAR} differs from audio synthesis as it focuses on spatial acoustic effects modelling.

\textbf{Time-series Prediction.} Predicting warping field in either time or frequency domain partially relates to time-series prediction. Existing deep neural network based time-series prediction methods can be divided into four main categories: Convolutional Neural Networks~(CNNs) based~\cite{zheng2014time, yang2015deep, wang2017time, foumani2021disjoint}, Recurrent Neural Networks based~\cite{dennis2019shallow, tang2016sequence, sutskever2014sequence}, Graph Neural Networks based~(GNNs)~\cite{covert2019temporal, jia2020graphsleepnet, ma2021deep, tang2022selfsupervised, zhang2021graph} and Transformer based~\cite{song2018attend, jin2021end, liu2021gated} methods. However, unlike existing works, warping field prediction in \emph{SPEAR} exhibits no causality and the warping field is estimated from receiver positions.
\vspace{-4mm}
\section{Receiver-to-Receiver Acoustic Neural Warping Field}
\vspace{-2mm}
\subsection{Problem Formulation}
\vspace{-2mm}
In a linear time-invariant~(LTI) enclosed 3D space $\mathcal{R}$, stationary sound sources are constantly emitting audio waveform. We use two receivers to record the audio at various discrete positions independently. At each time step, we temporally synchronize the two receivers before recording so that the two receivers are recording the same audio content. In addition to audio, we also record the two receivers' spatial position. Specifically, we have collected $N$-step paired receiver dataset $\{(\mathcal{A},\mathcal{P})=\{(x_{1,i}(t), p_{1,i}), (x_{2,i}(t), p_{2,i})\}_{i=1}^{N}\}$, $\mathcal{A}\in \mathbb{R}^{T\times N}$, $\mathcal{P}\in \mathbb{R}^{3\times N}$. The recorded audio $x_{1,i}(t)$ and $x_{2,i}(t)$ is the raw audio waveform in time domain~(both are of the same length $T$ sampled at the same sampling rate). $p_{1,i}$ and $p_{2,i}$ are the two receivers' spatial coordinate $[x_{k,i}^{\prime}, y_{k,i}^{\prime}, z_{k,i}^{\prime}]$ ($k\in \{1,2\}$). Our target is to learn a receiver-to-receiver acoustic neural warping field $\mathbfcal{F}$ from $(\mathcal{A},\mathcal{P})$, $\mathbfcal{F}\leftarrow (\mathcal{A},\mathcal{P})$, so that it can efficiently predict the spatial audio acoustic effects for any arbitrary target position $p_t$ by predicting a warping transform $\mathcal{W}_{p_r\rightarrow p_t}$ that warps audio recorded at another reference position $p_r$ to the target position $p_t$. The warped audio $\widehat{x_{p_r \rightarrow p_t}(t)}$ at position $p_t$ essentially accommodates the spatial acoustic effects belonging to $p_t$.

\begin{equation}
    \mathcal{W}_{p_r\rightarrow p_t} = \mathbfcal{F}_{\theta}(p_t, p_r);\ \ \widehat{X_{p_r\rightarrow p_t}(f)} = \mathcal{W}_{p_r\rightarrow p_t} \cdot X_{p_r}(f); \ \ \ p_t, p_r \notin \mathcal{P} 
\label{prbdef}
\end{equation}

Where $\theta$ is the trainable parameters of $\mathbfcal{F}$. $p_r$, $p_t$ are arbitrary positions in the 3D space $\mathcal{R}$. $\widehat{X_{p_r\rightarrow p_t}(f)}$ and $X_{p_r}(f)$ are discrete Fourier transform~(DFT) representation of the warped audio at the target position and recorded audio $x_{p_r}(t)$ at reference position represented in time domain, respectively. For example, if $x_{p_r}(t)$ has $T$ points, the $T$-point DFT result  $X_{p_r}(f)$ is a complex representation where both the real and imaginary part have $T$ data points. The learned $\mathcal{W}_{p_r\rightarrow p_t}$ is a complex representation of the same shape of $X_{p_r}(f)$. It is worth noting that, in our formulation, the warping transform is multiplication in frequency domain~(see Sec.~\ref{sec:math_background} for the proof), and the acoustic neural warping field is independent on audio content $\mathcal{A}$~(see Sec.~\ref{sec:warp_property} for more discussion). 

To optimize $\mathbfcal{F}_{\theta}$, we minimize the discrepancy $\mathcal{L}$ between the warped audio at the target position and corresponding recorded audio,
\vspace{-1mm}
\begin{equation}
    \mathbfcal{F}_{\theta} \leftarrow \argmin_{\theta} \ \mathcal{L}(\widehat{X_{p_1\rightarrow p_2}(f)}, X_{p_2}(f)), \ \ \ \forall p_1, p_2 \in \mathcal{P}
\end{equation}

\vspace{-3mm}
\subsection{Mathematical Backend of Receiver-to-Receiver Neural Warping Field}
\label{sec:math_background}

Before introducing \emph{SPEAR}, we need to answer two questions: 

\textbf{Q1:}. \textit{Does the proposed warping field suffice to model receiver-to-receiver spatial acoustic effects?}\\
\textbf{Q2:}. \textit{If so, is there any constraint on the audio source number and placement in the 3D space?}

We first consider the simplest case where there is just one audio source in the 3D space.

\textit{\textbf{Proposition 1:}} If the Linear Time-Invariant~(LTI) 3D space contains a single audio source, then receiver-to-receiver warping exists and is uniquely defined for any pair of receiver positions.

\textit{\textbf{Proof:}} Assume the audio source isotropically emits sound waveform $s(t)$ at a fixed position, the two receivers' recorded spatial audio in time domain at two different positions are $x_1(t)$ and $x_2(t)$, respectively. According to room acoustics~\cite{georoom_overview}, $x_1(t)$ and $x_2(t)$ are obtained by convolving with their respective impulse response RIR with the sound source $s(t)$,

\begin{equation}
    x_1(t) = s(t) \circledast h_1(t), \ \ \ x_2(t) = s(t) \circledast h_2(t)
\label{eqn:twosource_timedomain}
\end{equation}

where $h_1(t)$ and $h_2(t)$ are the two RIRs from the source $s(t)$ to receiver $x_1(t)$ and $x_2(t)$ respectively. $\circledast$ is the 1D convolution in time domain. According the Convolution theorem that time domain convolution equals to production in Frequency domain, we can rewrite Eqn.~(\ref{eqn:twosource_timedomain}) as,

\begin{equation}
    X_1(f) = S(f)\cdot H_1(f),\ \ \  X_2(f) = S(f)\cdot H_2(f)
\label{eqn:twosource_freqdomain}
\end{equation}

where $X(\cdot)$, $S(\cdot)$ and $H(\cdot)$ are the Fourier transform of receiver recorded audio, source audio and RIR, respectively. Based on Eqn.~(\ref{eqn:twosource_freqdomain}), we can further get,

\begin{equation}
      X_2(f) = X_1(f)\cdot \frac{H_2(f)}{H_1(f)}, \ \ X_1(f) = X_2(f)\cdot \frac{H_1(f)}{H_2(f)}
\label{eqn:twosource_warpfield}
\end{equation}

Let $\mathcal{W}_{1\rightarrow 2} = \frac{H_2(f)}{H_1(f)}$~(or $\mathcal{W}_{2\rightarrow 1} = \frac{H_1(f)}{H_2(f)}$), we can conclude that: 1) the receiver-to-receiver warping universally exists and 2) is uniquely defined for any pair of receiver positions, 3) it relies on the 3D space and is independent on audio source.

\textit{\textbf{Proposition 2}}: If the more than one audio sources are placed in the 3D space, the receiver-to-receiver warping field existence is not guaranteed.

\textit{\textbf{Proof:}} Assume $K$~($K>1$) audio sources are placed in the 3D space, based on the superposition property in room acoustics~\cite{georoom_overview}, one receiver recorded audio~(\textit{e.g.}, $x_1(t)$) can be expressed as, 

\begin{equation}
x_1(t) = s_1(t) \circledast h_{1,1}(t) + s_2(t) \circledast h_{2,1}(t) + \cdots + s_K(t) \circledast h_{K,1}(t)
\label{eqn:multisource_timedomain}
\end{equation}

where $h_{k,l}(t)$ indicates the RIR from the $k$-th source to the $l$-th receiver, By extending Eqn.~(\ref{eqn:multisource_timedomain}) to $M$ receivers and further applying Fourier transform, we can get,

\begin{equation}
    \left[\begin{array}{c}
         X_1(f)  \\
         X_2(f) \\
         \cdots \\
         X_M(f)
    \end{array}\right] = \left[\begin{array}{cccc}
         H_{1,1}(f) &  H_{1,2}(f) & \cdots & H_{1,K}(f)\\
         H_{2,1}(f) &  H_{2,2}(f) & \cdots & H_{2,K}(f) \\
         \cdots & \cdots & \cdots & \cdots \\
         H_{M,1}(f) &  H_{M,2}(f) & \cdots & H_{M,K}(f) \\
    \end{array}\right] \cdot     \left[\begin{array}{c}
         S_1(f)  \\
         S_2(f) \\
         \cdots \\
         S_K(f)
    \end{array}\right]
\label{eqn:eqn:multisource_freqdomain}
\end{equation}

For brevity, we can rewrite Eqn.~(\ref{eqn:eqn:multisource_freqdomain}) as $X = H\cdot S$. Since we have no knowledge of audio sources $S$, we can treat Eqn.~(\ref{eqn:eqn:multisource_freqdomain}) as multivariate polynomial linear function for audio sources $S$. To ensure a unique solution for $S$, two conditions have to be satisfied: first, the determinant of the coefficient matrix $H$ must be non-zero, det($H$) $\neq 0$. Second, the rank of coefficient matrix $H$ must be equal to the rank of the augmented matrix $[H|X]$, rank($H$) $=$ rank($[H|X]$). That is to say, special receiver placement is required to deterministically represent audio sources by receivers. Moreover, even if we can deterministically represent audio source by receivers, $S = H^{-1}\cdot X$, we can hardly predict one receiver's spatial acoustic effects by warping from another single receiver because one receiver's spatial acoustic effects, in multiple audio sources case, usually depend on multiple other receivers. We empirically show one such example in Appendix~\ref{appdix_warpfield_twosrc}.

\vspace{-2mm}
\subsection{LTI Receiver-to-Receiver Warping Field Physical Principle}
\label{sec:warp_property}
\vspace{-1mm}

We present three room acoustics physical principles~\citep{room_acoustics,Rayleightos} that will guide \emph{SPEAR} design. 

\textit{Principle 1,} \textbf{Globality:} Unlike a normal RGB image just captures a localized area, a receiver recorded spatial audio relates to the whole 3D space. Originating from the source position, the spatial audio propagates in a complex way that incorporates reflection, diffraction and absorption before reaching to the receiver position, resulting in the interaction with almost the whole 3D space before reaching to the receiver. Consequently, the final receiver recorded audio is influenced by the whole 3D space.

\textit{Principle 2,} \textbf{Order Awareness}: The Order Awareness principle states that \emph{SPEAR} should account for the specific order of the input two receivers.  In essence, the learned warping field varies when the order of the two receivers is swapped. This can be readily proved by Eqn.~(\ref{eqn:twosource_warpfield}), since $\mathcal{W}_{1\rightarrow 2} = \frac{H_2(f)}{H_1(f)}$, $\mathcal{W}_{2\rightarrow 1} = \frac{H_1(f)}{H_2(f)}$, $\mathcal{W}_{1\rightarrow 2} \neq \mathcal{W}_{2\rightarrow 1}$.

\textit{Principle 3,} \textbf{Audio-Content Agnostic}: This principle asserts that the receiver-to-receiver warping field is an inherent characteristic of the 3D space, affected by neither the presence or absence of audio within the 3D space nor the specific class of audio.

\begin{figure*}[t]
    \centering
    \includegraphics[width=0.98\linewidth]{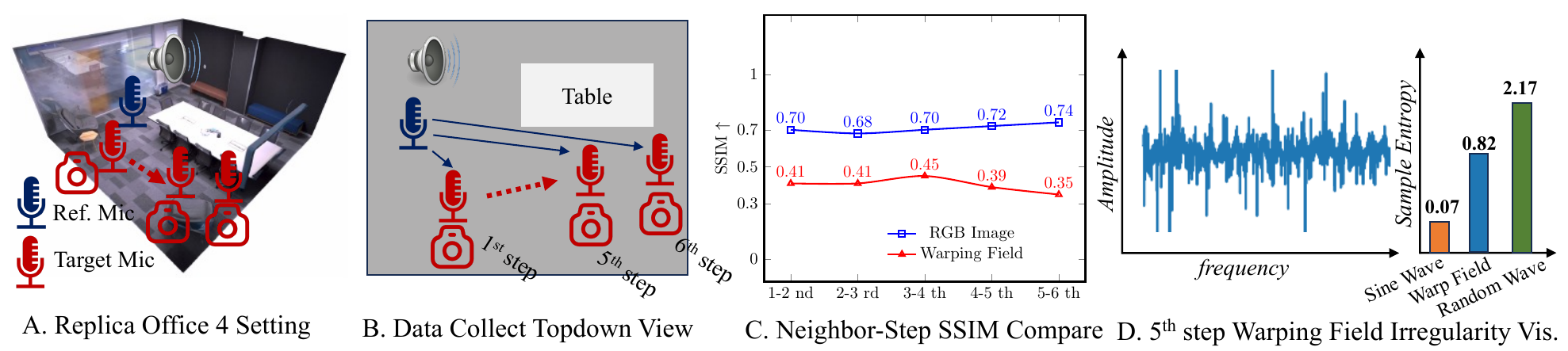}
    \caption{Two challenges in \emph{SPEAR} learning: \textbf{Position-Sensitivity} and \textbf{Irregularity}. The position-sensitivity is represented by much lower structural similarity index~(SSIM) of two neighboring-step warping fields than the two RGB images~(sub-fig.~C). The warping field irregularity is represented by both warping field visualization in frequency domain~(real part) and much higher sample entropy score than regular sine wave~(and just half of random waveform)~(sub-fig. D).}
    \label{fig:warp_irregu_vis}
\end{figure*}

\vspace{-2mm}
\subsection{Position-Sensitivity and Irregularity of Receiver-to-Receiver Neural Warping Field}
\label{sec:warp_challenge}
\vspace{-1mm}

The complex behavior of spatial audio propagation in an enclosed 3D space often results in significantly different spatial acoustic effects even for audio recorded at neighboring positions. This position-sensitivity becomes even more pronounced in the receiver-to-receiver warping field, where even a small receiver position can lead to substantial warping field variation. As is shown in Fig.~\ref{fig:warp_irregu_vis}, we compare the visual differences and warping field variations caused by small receiver position change. Using the 3D space \texttt{Office 4} from the Replica dataset~\cite{replica19arxiv} and the SoundSpaces 2.0 simulator~\cite{chen22soundspaces2}, we place a stationary audio source and a reference receiver. Next, we ask the robot carrying a pinhole camera and a receiver to walk straightforward with step size $0.3$~m in the vicinity of the reference receiver. At each target receiver position~(crimson color in Fig.~\ref{fig:warp_irregu_vis}), we capture RGB images and compute the corresponding warping fields. We then adopt structural similarity index~(SSIM)~\cite{SSIM-metric} to measure the visual and warping field differences between two neighboring positions. The results, shown in Fig.~\ref{fig:warp_irregu_vis}, clearly indicate that a 0.3-meter position change results in a significantly more pronounced warping field variation (much lower SSIM score) compared to the RGB images. To show warping field irregularity, we visualize the $5$-th step warping field real part in frequency domain in Fig.~\ref{fig:warp_irregu_vis} D, from which we can see the warping field is highly irregular and thus exhibits higher sample entropy score~\cite{sample_entropy}\footnote{The higher of sample entropy score, the more irregular is the series. To highlight warping field irregularity, we compare its sample entropy score with totally regular sine wave and random wave.} than regular sine wave.

\vspace{-3mm}
\subsection{SPEAR Neural Network Introduction}
\vspace{-2mm}

\begin{wrapfigure}{r}{0.40\textwidth}
        \centering
    \includegraphics[width=0.40\textwidth]{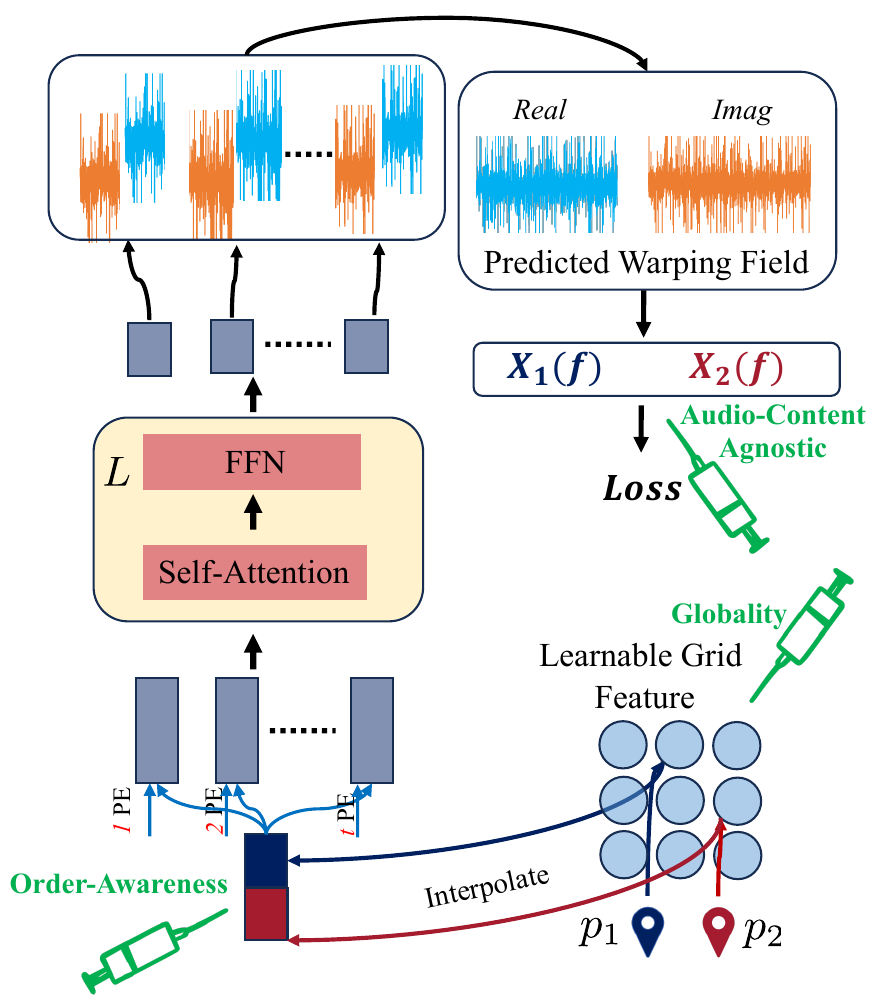}
    \caption{\emph{SPEAR} network visualization.}
    \label{fig:spear_pipeline}
\end{wrapfigure}

The way we design \emph{SPEAR} neural network is guided by Sec.~\ref{sec:warp_property} and Sec.~\ref{sec:warp_challenge}. Specifically, \emph{SPEAR} takes as input two receivers' positions~(one reference position and the other target position) and outputs the corresponding warping field that warps the spatial acoustic effects at the reference position to the target position~(\textbf{Audio-Content Agnostic} principle applies). We build \emph{SPEAR} on top of Transformer architecture~\cite{att_all_need} and predict the warping field in frequency domain, in which the warping field is jointly represented by a real and imaginary series. Predicting the warping field in frequency domain results in both faster processing and better generation quality, as shown in the ablation experiment~\ref{sec:ablation}. 
To accommodate the \textbf{Globality} principle and tackle the position-sensitivity challenge, we construct a learnable grid feature spanning to the whole 3D space horizontally, each cell of the grid thus corresponds to one particular physical position in the 3D space. 
The two input positions' features are extracted from the grid feature by bilinear interpolation.
Concatenating the two interpolated features in order~(\textbf{Order-Awareness} principle satisfied) gets the input position-pair's representation. Each token then combines the input positions' representation and token index position encoding as the initial token representation. 
After 12 transformer blocks, the final learned token representation is further fed a prediction head to predict their corresponding real/imaginary warping field. We visualize the neural architecture in Fig.~\ref{fig:spear_pipeline}.

\textbf{Train and Inference.} 
To ensure the audio source covers the whole frequency range of the warping field, we choose Sine Sweep audio with frequency range [0-16]~kHz. As only one audio source presents in the 3D space, we can obtain the ground truth warping field by dividing the target position received audio by source position received audio~(see Eqn.~(\ref{eqn:twosource_warpfield})). We thus jointly train \emph{SPEAR} with group truth warping field and two receivers' recorded audio. The original warping field length we predict is 32~k. Due to the warping field symmetry~(DFT conjugate symmetry), we just predict half warping field which are 16~k points. For ground truth warping field supervision, we combine both L1 and L2 loss. For the two receivers recorded audio supervision, we adopt spectral convergence loss~\cite{spectral_convergence}. More detailed discussion is in supplementary material.
\vspace{-3mm}
\section{Experiment}
\vspace{-3mm}
\subsection{Experiment Configuration} \label{sec:exp_config}
\vspace{-2mm}

\textbf{Dataset.} We evaluate \emph{SPEAR} on three datasets across different domains. 

1. \textbf{Synthetic Dataset.} We adopt Pyroomacoustics~\cite{pyroomacoustic} to simulate a large shoebox-like 3D space of size $[5m \times 3m \times 4m]$. All the receivers are placed on the same height, the audio source is placed at position $[2m, 2m, 2m]$. We simulate 3000 receivers positions for training and another 300 receivers for test, by ensuring the reference-target receiver pair position in the test set is significantly different from those in training set. 

2. \textbf{Photo-realistic Data.} To show our model is capable of predicting the warp field in a more complicated and photo-realistic environment, we employ Replica \texttt{Office 0} and \texttt{Office 4} 3D space~\cite{replica19arxiv} to simulate the spatial audio. In \texttt{Office 0} and \texttt{Office 4}, we simulate 4000 receivers and 8000 receivers for training, respectively. Both scenes have 500 receiver positions for test.

3. \textbf{Real-World Dataset.} We further test on the real-world classroom 3D space~\cite{C4DM_real_rir_dataset} of size $(9~m\times 8~m \times 3.5~m)$. The audio source is placed on the platform in the classroom. A grid of $[10\times 13]$ receivers are placed on the on ground with spatial resolution $0.5~m$. 106 of the audio positions are used for training, and the rest 24 positions are reserved for test. 

\begin{wrapfigure}{r}{0.5\textwidth}
    \centering
    \includegraphics[width=0.48\textwidth]{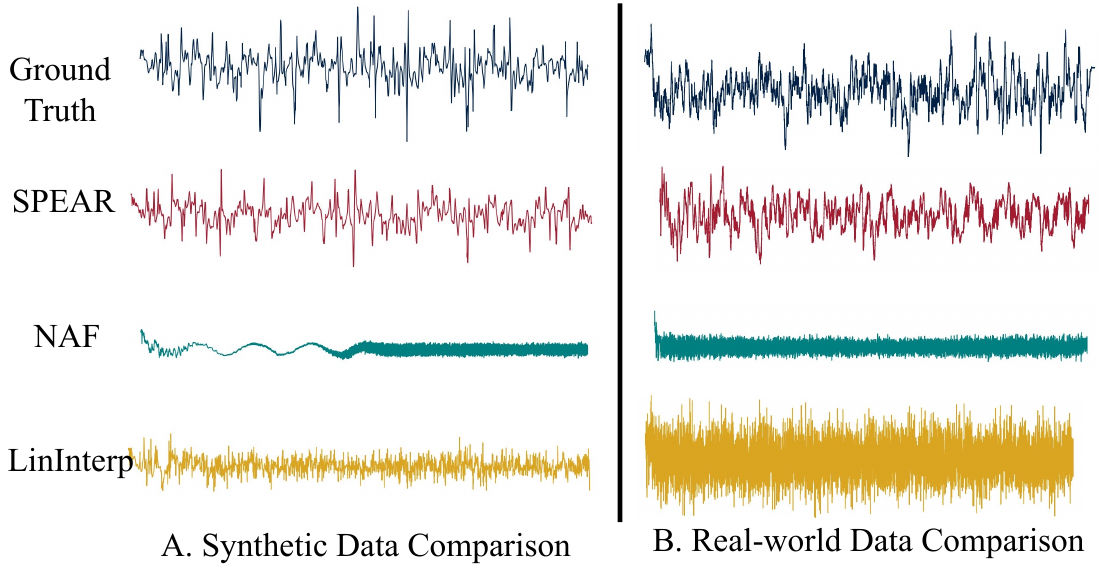} 
    \caption{\small Learned warping field visualization on synthetic dataset~(A) and real-world dataset~(B).}
    \label{fig:warpfield_vis}
\end{wrapfigure}

\textbf{Evaluation Metrics}. To quantitatively evaluate the warping field, we adopt: 1. \textbf{MSE}, mean square error between ground truth and \emph{SPEAR} predicted warping field in frequency domain~(average between real and imaginary part). 2.~\textbf{SDR}~(signal-to-distortion ratio). Following~\cite{richard2022deepimpulse}, we report SDR to assess the fidelity of predicted warping field. 3.~\textbf{PSNR}~(Peak Signal-to-Noise Ratio~\cite{PSNR}) and 4.~\textbf{SSIM}~(structural similarity index measure~\cite{SSIM-metric}) to quantify the quality of the predicted warping field. We also introduce human-perceptual metric to provide insight into human perceptual quality of the predicted warping field. Specifically, we choose five speeches from VCTK~\cite{yamagishi2019vctk} dataset and warp them to the target position with the learned warping field, then compute the 5.~\textbf{PSESQ}~(perceptual evaluation of speech quality~\cite{PESQ_eval}) score for the warped speech and ground truth recorded speech.

\begin{table*}[t]
    \centering
    \scriptsize
    \caption{\small Quantitative result on three datasets. MSE~($\downarrow$), SDR~($\uparrow$), PSNR~($\uparrow$), SSIM~($\uparrow$), PSEQ~($\uparrow$).}
    \begin{tabular}{l|p{0.3cm}<{\centering}p{0.3cm}<{\centering}p{0.4cm}<{\centering}p{0.4cm}<{\centering}|p{0.4cm}<{\centering}|p{0.3cm}<{\centering}p{0.3cm}<{\centering}p{0.4cm}<{\centering}p{0.4cm}<{\centering}|p{0.4cm}<{\centering}|p{0.3cm}<{\centering}p{0.3cm}<{\centering}p{0.4cm}<{\centering}p{0.4cm}<{\centering}|p{0.4cm}<{\centering}}
    \hline
    \multirow{2}{*}{Method}&\multicolumn{5}{c|}{Synthetic Data} & \multicolumn{5}{c|}{Photo-Realistic Data} & \multicolumn{5}{c}{Real-World Data} \\
    \cline{2-16}
    & SDR & MSE & PSNR & SSIM & PSEQ & SDR & MSE & PSNR & SSIM & PSEQ & SDR & MSE & PSNR & SSIM & PSEQ\\
    \hline
    LinInterp  &  $-$0.37 &  1.40 &  14.58 & 0.85 & 1.65 & $-$0.94 & 1.44 & \cellcolor{topcolor}\textbf{14.71} & 0.63 & 2.16 &  $-$1.02 & 1.20 & \cellcolor{topcolor} \textbf{14.20} &  0.49 &  1.27\\
    NNeigh  & $-$2.71 & 2.44 & 15.21 & 0.83 & 1.62 & $-$2.87 & 2.22 & 12.13 & 0.64 & 1.89 & $-$3.78 & 2.27 & 12.13 & 0.44 &  1.30 \\
    NAF~\cite{NAF}   &  0.13 &  1.24  & 13.08 & 0.86 & 1.77 &   0.07 & 1.13  & 14.21 & 0.73 & 1.92 &  $-$0.24 &  1.37  & 13.01 & 0.33 &  1.30\\
    \hline
     \emph{SPEAR}  & \cellcolor{topcolor}\textbf{1.50} & \cellcolor{topcolor}\textbf{0.92} & \cellcolor{topcolor}\textbf{15.81} & \cellcolor{topcolor}\textbf{0.87} & \cellcolor{topcolor}\textbf{2.47}  & \cellcolor{topcolor}\textbf{0.66} & \cellcolor{topcolor}\textbf{1.03} & 14.57 & \cellcolor{topcolor}\textbf{0.75} & \cellcolor{topcolor}\textbf{2.18}  & \cellcolor{topcolor}\textbf{$-$0.24} & \cellcolor{topcolor}\textbf{1.04} & 14.07 & \cellcolor{topcolor}\textbf{0.75} & \cellcolor{topcolor}\textbf{1.45}\\
    \hline
    \end{tabular}
    \label{tab:all_result}
    \vspace{-5mm}
\end{table*}
\textbf{Comparison Methods} Given the novel problem setting of \emph{SPEAR}, currently there is no existing method that directly applies to our problem. Existing RIR-based spatial acoustic effects modelling methods~\cite{IR-GAN,TS-RIR,improved_RIR} vary significantly in the way they exploit RIR data and the amount of prior room scene acoustic properties knowledge to train their model, we thus find it difficult to modify them to fit our setting. For meaningful comparison, we compare \emph{SPEAR} with one source-to-receiver neural acoustic field learning method~(NAF~\cite{NAF}) and two other interpolation methods.

1. \textbf{NAF}~\cite{NAF}. NAF learns continuous source-to-receiver RIR field by assuming access to massive RIR data. We modify it to accept two positions as input and output the warping field in frequency domain. 

2. \textbf{LinInterp}: Neighboring Linear-Interpolation. For each receiver-pair in test set, we retrieve top-25 warping field data from the training set whose position are closest to the receiver pair~(the position distance is defined by the the sum of reference position shift and target position shift). Then we average the 25 warping field to get the linear-interpolated warping field.

3. \textbf{NNeigh}: Nearest Neighbor method searches for the warp field, where the reference and target receiver positions are closest to the input test position pair. This is equivalent to retrieving the top-1 closest warping field from the training set. 

\vspace{-2mm}
\subsection{Experimental Result}
\vspace{-2mm}
The quantitative result on the three datasets is given in Table~\ref{tab:all_result}, from which we can see that \emph{SPEAR} outperforms all the three comparing methods by a large margin. Among all the five metrics, \emph{SPEAR} maintains as the best-performing method~(except for one PSNR metric). Moreover, \emph{SPEAR} outperforms the second-best method NAF~\cite{NAF} significantly on metrics like SDR and PESQ.

We also provide qualitative comparison of the predicted warping field in Fig.~\ref{fig:warpfield_vis}, in which we visualize the warping field~(real-part) on Synthetic dataset~(A) and real-world dataset~(B) obtained by all methods. From this figure, we can clearly observe that 1) while the ground truth warping field exhibits high irregularity~(see Sec.~\ref{sec:warp_challenge}), \emph{SPEAR} is capable of learning the irregularity pattern from the input position pair; 2) The two other methods~(NAF~\cite{NAF} and LinInterp) failed to tackle the irregularity challenge. NAF~\cite{NAF} tends to predict close warping field values. The higher of the frequency, the easier it tends to predict the same value~(see Fig.~\ref{fig:warpfield_vis} A, the left and right part comparison of NAF).

\vspace{-2mm}
\subsection{Ablations} \label{sec:ablation}
\vspace{-2mm}

\begin{figure*}
\begin{minipage}[b]{.734\linewidth}
\centering
\includegraphics[width=0.98\linewidth]{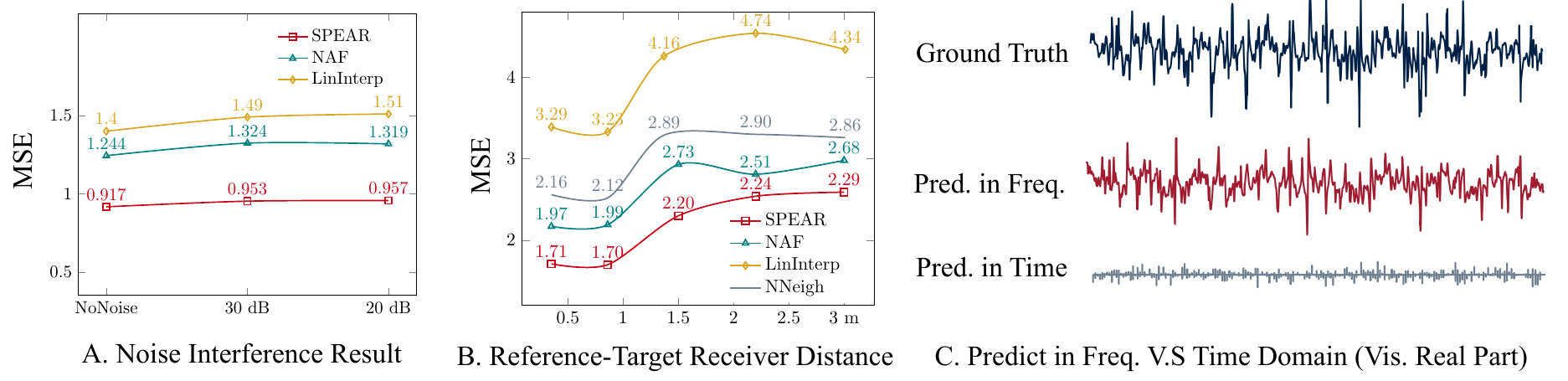}
\caption{\small Ablation Study on noise interference~(A), reference-target receiver distance~(B) and Prediction in frequency domain and time domain.}\label{fig:ablation_study}
\end{minipage}\quad
\begin{minipage}[b]{.24\linewidth}
\centering\resizebox{.99\linewidth}{!}{\renewcommand\arraystretch{2.28}
\begin{tabular}{l|cccc}
\hline
Metrics & Chirp & Engine & Person & Siren \\
\hline
PSNR & 15.815 & 15.814  & 15.740 & 15.780 \\
SDR  & 1.503 & 1.503 & 1.435 & 1.478 \\
SSIM & 0.869 & 0.868 & 0.848 & 0.849  \\
MSE  & 0.917  & 0.917 & 0.933 & 0.925  \\
\hline
\end{tabular}}
\captionof{table}{\small Ablation study on audio-content agnostic.}
\label{tab:source_agnostic}
\end{minipage}
\end{figure*}

We run all ablation studies on Synthetic dataset created by Pyroomacoustics~\cite{pyroomacoustic}.

\textbf{Warping Field Sensitivity to Noise.} We add two ambient Gaussian noises~(measured by SNR, 20~dB and 30~dB) to test model's performance under noise interference. As is shown in Fig.~\ref{fig:ablation_study}~A, we can observe that while all three comparing methods have seen performance drop~(increased MSE metric) as more noise is involved, \emph{SPEAR} maintains as the best-performing method and outperforms the other two methods~(NAF and LinInterp) by a large margin under all noise interference.  

\textbf{Warping Field Accuracy w.r.t. Reference-Target Receiver Distance.} We further test \emph{SPEAR}'s capability in predicting spatial acoustic effects for spatially distance target receivers. To this end, we compute models' performance variation w.r.t. reference-target receivers' distance change~($0.5~m-3.0~m$) and show the result in Fig.~\ref{fig:ablation_study}~B. We can observe from this figure that: 1) The increased reference-target receiver's distance leads to performance drop~(increasing MSE) for all methods, which is expected as the spatial acoustic effects drastically change when the receiver position gets changed dramatically. 2) \emph{SPEAR} still outperforms the other comparing methods, showing its advantage in predicting warping field for further target receivers.

\textbf{Warping Field Prediction in Frequency V.S. Time domain.} In \emph{SPEAR}, we propose to predict warping field in frequency domain for computation efficiency concern. Out of the computation efficiency concern, we further want to figure out the performance with predicting in time domain. As depicted in Fig.~\ref{fig:ablation_study}~C, we can observe that predicting warping field in time domain leads to significant performance drop as the model tends to predict all-zero warping field.

\textbf{Audio-Content Agnostic Verification.} Since all models are trained with Sine Sweep audio, we want to know if the \textit{Audio-Content Agnostic} principle truly satisfies. To this end, we evaluate \emph{SPEAR} with another three audios: engine, siren and human vocalization. As is shown in Table~\ref{tab:source_agnostic}, we can verify that \emph{SPEAR} is agnostic to audio content and can be applied to arbitrary audio class.

More experimental result and discussion are provided in supplementary material.

\section{Conclusion and Limitation Discussions}
\vspace{-3mm}
We introduce a novel receiver-to-receiver spatial acoustic effects prediction framework that can be trained in a data much more readily accessible way. We theoretically prove the existence and universality of such warping field if there is only one audio source. When there are multiple sources, we can apply source separation method~\cite{source_separation_icassp} first to separate the sources first. There are two limitations that remain to be resolved in the future. The first is that we still require dense sampling to get reasonably good performance. Second, we assume all receivers' position lies on the same horizontal plane. More in-depth investigation is needed to remove this constraint and allow receiver placement in the whole 3D space.

\clearpage

\bibliography{egbib}  
\newpage
\renewcommand{\thetable}{\Roman{table}}
\renewcommand{\thefigure}{\Roman{figure}}
\renewcommand\thesection{\Alph {section}}

\appendix
\section*{Appendix}
\setcounter{section}{0}
\setcounter{figure}{0}
\setcounter{table}{0}

\section{Receiver-to-Receiver Warping Field Existence Discussion}
\label{appdix_warpfield_twosrc}

\begin{figure*}[h]
    \centering
    \includegraphics[width=0.98\linewidth]{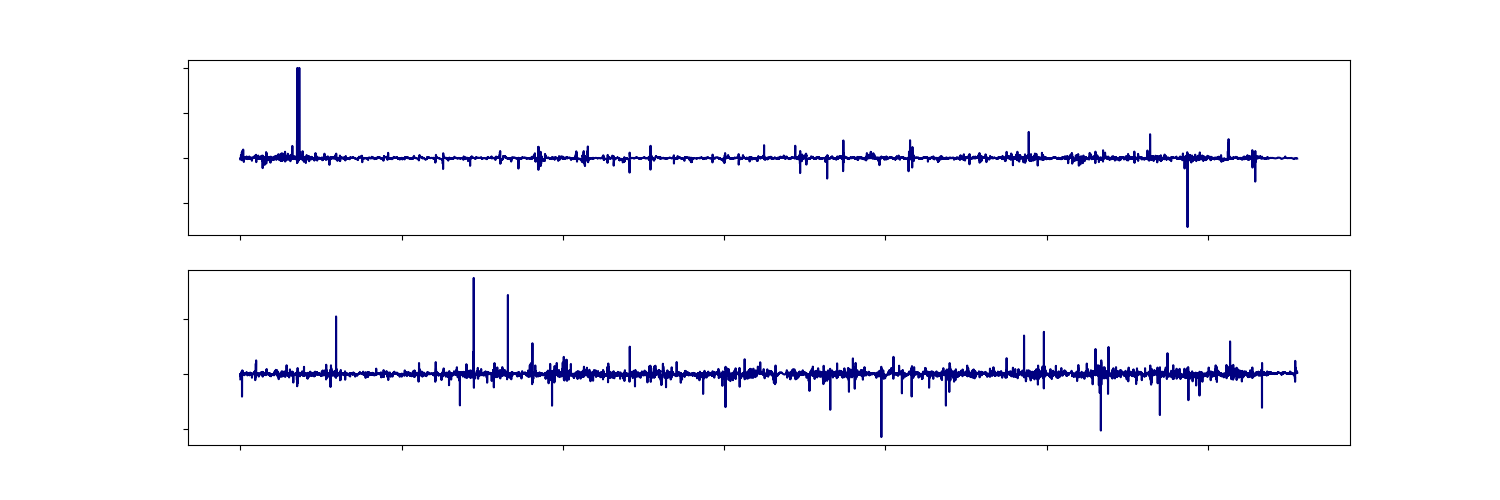}
    \caption{Warping field visualization in frequency domain~(real part). Top: warping field with two audio sources \texttt{Engine} and \texttt{Footstep} two audio sources. Bottom: warping field with the two audio sources at the same positions but the \texttt{Footstep} sources is replaced by \texttt{Telephone Ring}.}
    \label{fig:2-source-warp-field}
\end{figure*}

In this section, we empirically show that the receiver-to-receiver neural warping field may not exist if more than one audio sources present in the 3D space. To this end, we depend on Pyroomacoustics~\cite{pyroomacoustic} simulator to simulate a shoe-box like 3D space of shape $[5\times 5 \times 5]$ meters. Two audio sources are placed at coordinate A $[1, 1, 1]$ meters and B $[4, 4, 4]$ meters, respectively. Two receivers are accordingly placed at coordinate $[1,1,2]$ meters and $[4, 4, 3]$ meters respectively. In the fist simulation, we place \texttt{Engine} audio at source position A and \texttt{Footstep} audio at source position B. In the second simulation, we just replace the \texttt{Footstep} audio at position B with \texttt{Telephone Ring}. With the two receivers recorded audio, we depend on Eqn.~(\ref{eqn:twosource_warpfield}) to calculate the warping field from the reference receiver at $[1,1,2]$ to the target receiver at $[4,4,3]$. The computed two warping fields are shown Fig.~\ref{fig:2-source-warp-field}, we clearly see that the two warping fields are significantly different. We thus can conclude that the receiver-to-receiver neural warping field is no longer solely dependent on receiver positions~(so the \textbf{Audio-Content Agnostic} principle does not apply). When more than audio sources present in the 3D space, the warping field proposed in this work may not exist.

\section{Discussion on Acoustic Neural Warping Field Visualization}
\begin{figure}[h]
    \centering
    \hspace*{-0.5in}
    \includegraphics[width=1.3\textwidth]{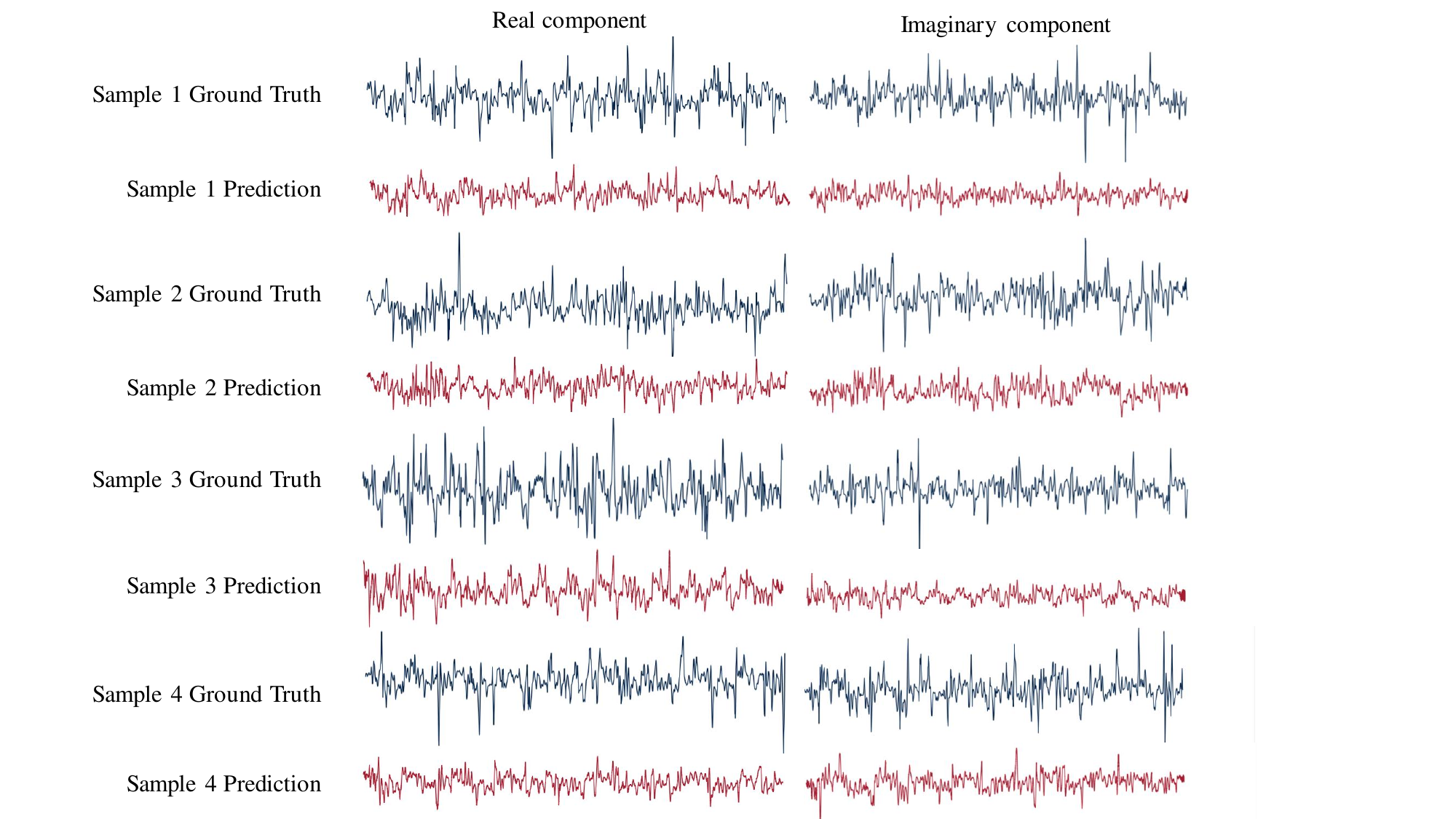}
    \caption{Visualization of real and imaginary component of the ground truth warp field in the synthetic dataset.}
    \label{fig:imaginary-part}
\end{figure}

The acoustic neural field is represented in frequency domain, each of which in our case contains a real and imaginary one-dimensional data vector. In the main paper, we just visualize the real part due to the space limit. Here, we provide another five real/imaginary warping fields visualization in Fig.~\ref{fig:imaginary-part}. From this figure, we can clearly see that our proposed framework~\emph{SPEAR} is capable of handling the warping field irregularity property.

\section{Failure Case Visualization}

\begin{figure}
    \centering
    \hspace*{-0.5in}
    \includegraphics[width=1.2\textwidth]{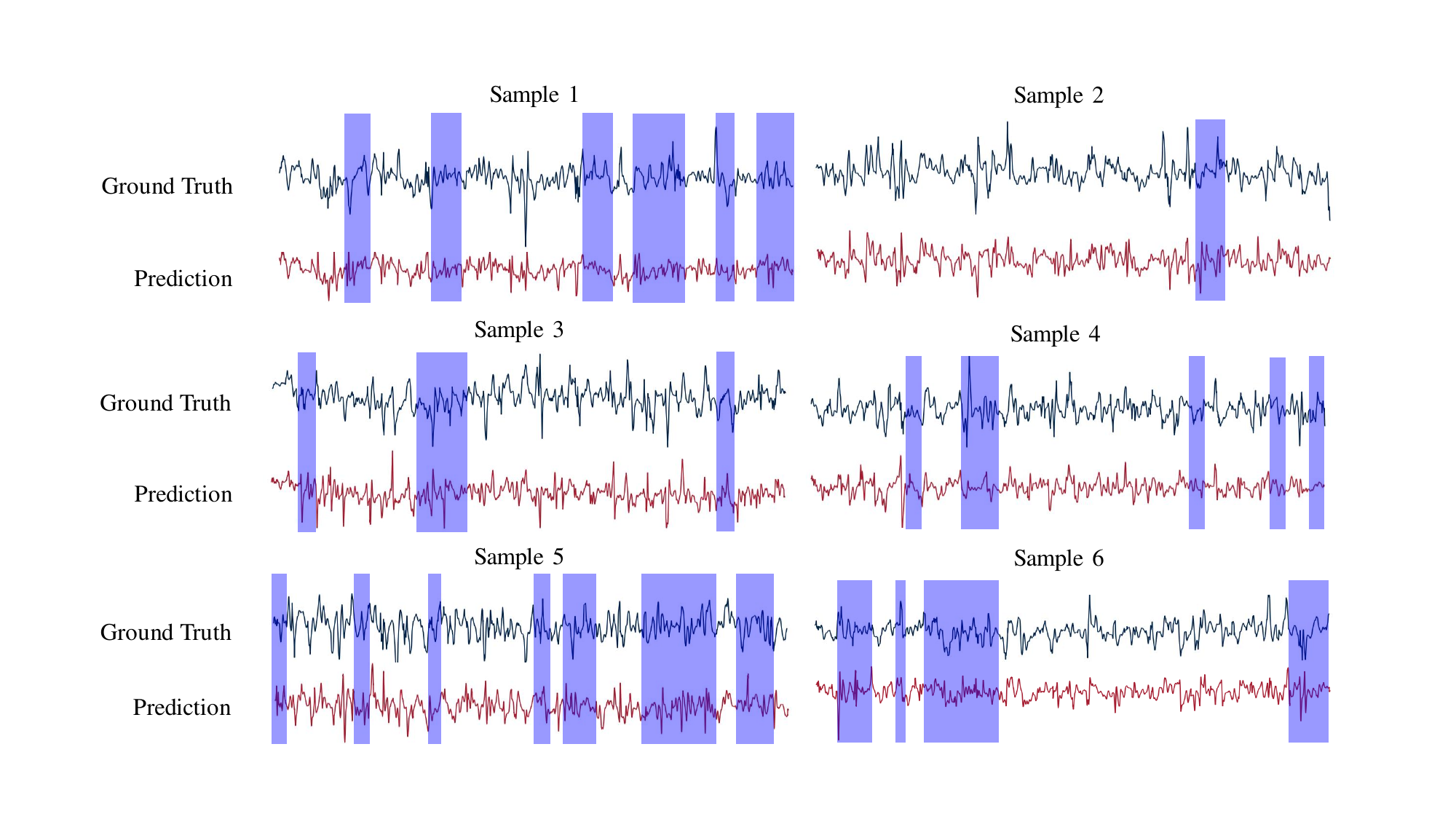}
    \caption{Examples of failure cases of SPEAR model on synthetic data. Regions with significant miss-match between the ground truth and the predicted warping field pattern are shaded in blue.}
    \label{fig:fail-case}
\end{figure}

\begin{figure}
    \centering
    \hspace*{-0.5in}
    \includegraphics[width=1.1\textwidth]{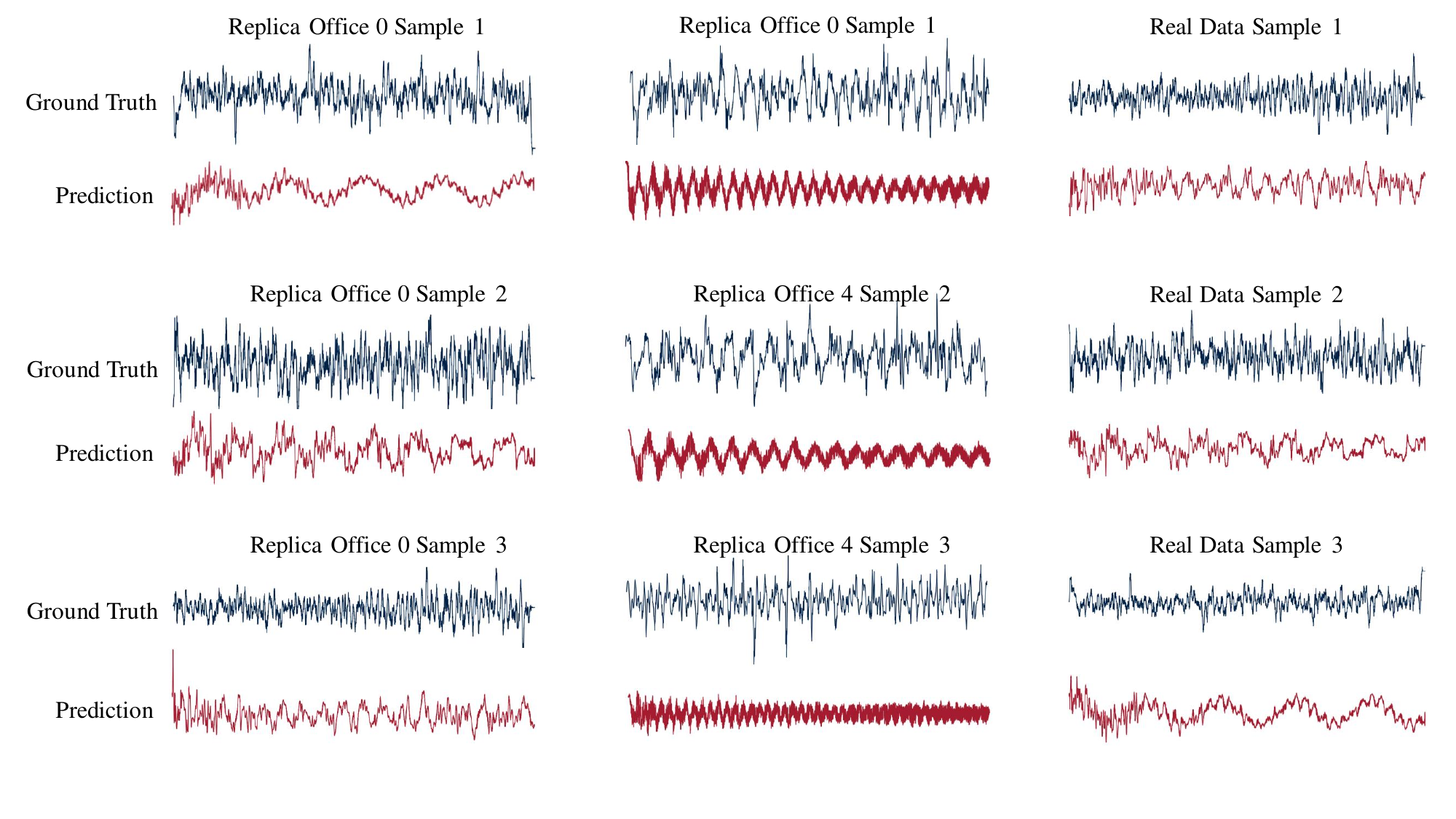}
    \caption{Failure case visualization of \emph{SPEAR} model on both Photo-realistic and Real-world Dataset.}
    \label{fig:replica-real-fail-case}
\end{figure}

During the experiment process, we find all methods inevitably give failure case warping field predictions. We visualize part of some failure cases predicted by \emph{SPEAR} on both synthetic data~(Fig.~\ref{fig:fail-case}), photo-realistic and real-world dataset~(Fig.~\ref{fig:replica-real-fail-case}). As we discussed in the main paper, the position-sensitivity and irregularity pose challenges in the warping field prediction. We hope these failure cases will attract more investigation into this research problem.

\section{Training Detail Presentation}
\label{sec:training-detail}

\subsection{Ground Truth Warping Field Acquirement}

\begin{figure}[t]
    \centering
    \hspace*{-0.25in}
    \includegraphics[width=1.1\textwidth]{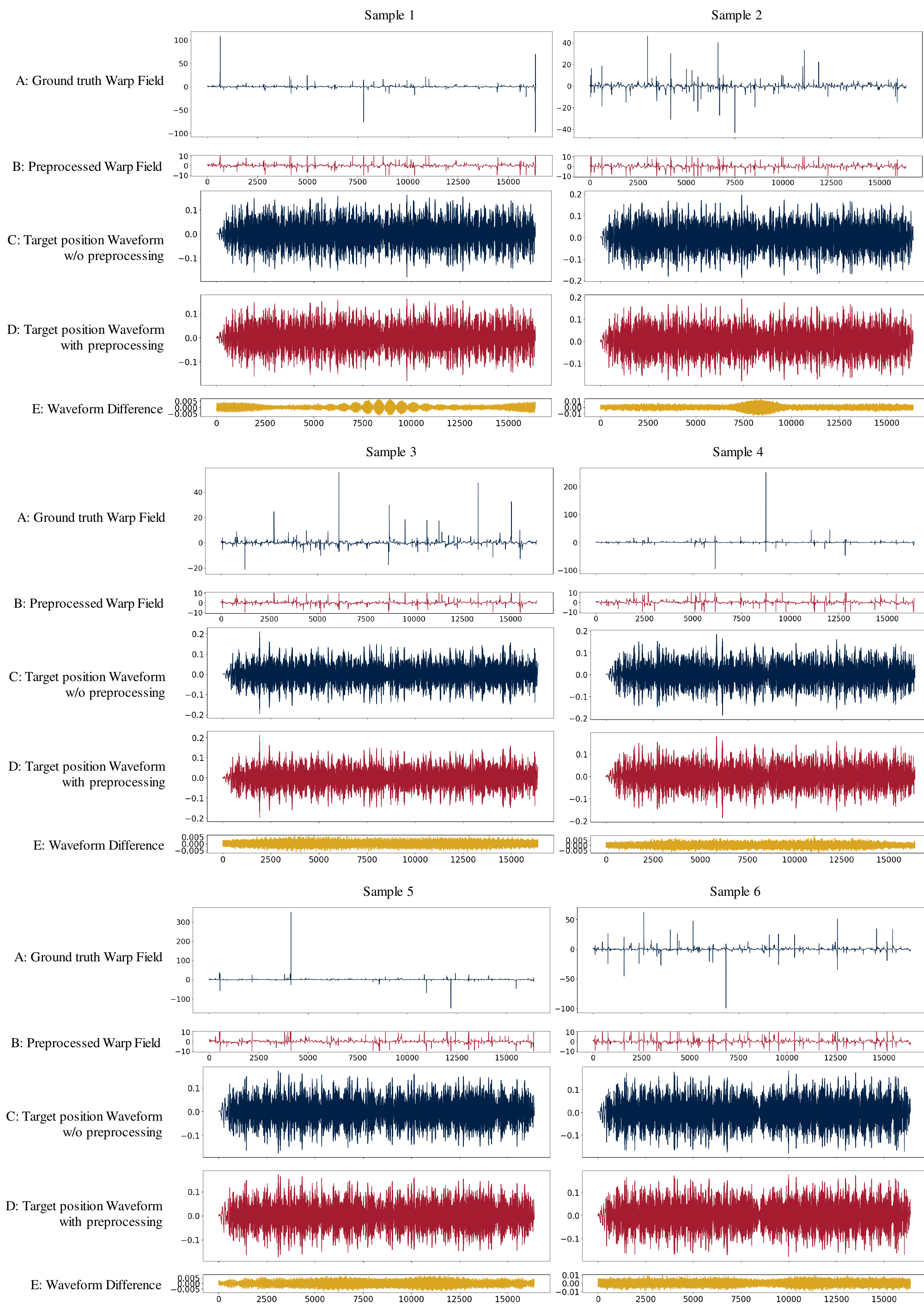}
    \caption{Visualization of the effect of preprocessing Warping field on the generated audio waveform. In plot A C, we show the warping field and waveform of the target position without warping-field preprocessing. In plot B D, we show the pre-processed warping field and the waveform at the target position after applying the preprocessed warping field. Plot E shows the difference between the two waveforms at the target position.}
    \label{fig:preprocessing}
\end{figure}

We obtain the ground truth by dividing the target receiver audio by the reference receiver audio in frequency domain. This division operator sometimes leads to \texttt{NaN} value or abnormally large value~($>100$, when the denominator is close to zero) in the obtained warping field~(see the ground truth warping field visualization in Fig.~\ref{fig:preprocessing}), resulting in the difficulty of accurately warping field learning. To handle this dilemma, we make two adjustments: first, replace \texttt{NaN} value with zero so that the whole neural network is trainable with the warping field prediction loss, which was \texttt{NaN} without the replacement.
Second, clipping all warping field values to lie within $[-10, 10]$. The reason for the \texttt{clip} operation is two-fold: the abnormally large value easily allures the whole neural network to be trapped in predicting those abnormally large values, thus ignoring predicting the warping field with normal values; we further empirically verify in Fig.~\ref{fig:preprocessing} that \texttt{clip} operation gives subtle difference in the warped target audio.

\subsection{Training Configuration Presentation}

We adopt AdamW optimizer~\cite{adamW_optimizer} for training on all datasets. On the synthetic dataset, the model requires approximately 10000 epochs to converge, which takes around 11 hours on a single A10 GPU. We set the learning rate of the learnable grid feature to 1e-5, and the rest learnable parameters' learning rates to 1e-4. Using a smaller learning rate for the grid features improves the model training stability. Since the model predictions rely solely on the grid feature extracted, changes in grid feature can result in significant differences in model prediction. Therefore, setting a lower learning rate for the grid features prevents the model prediction from changing abruptly, and thus improves training stability. 

Though our model has larger parameter size, the inference time is smaller than the NAF baseline. As shown in Tab.~\ref{tab:memory-time}, our model has more than ten times larger parameter size than the NAF model, but inference speed is around ten times faster than the NAF model.

\begin{table}[]
    \centering
    \begin{tabular}{c|cc}
    \hline
    \textbf{Method} & Inf. Time & Param. Num. \\
    \hline
    NAF~\cite{NAF} & 0.13~s & 1.61~M \\
    \emph{SPEAR} & 0.0182~s & 27.26~M \\
    \hline
    \end{tabular}
    \caption{\small Model param. and Inference Time comparison. The inference time is the average of 1000 independent inferences with batch size 32 on a single A10 GPU.}
    \label{tab:memory-time}
\end{table}

\subsection{Data Sampling Strategy}

For all three types of datasets, we construct train and test datasets by first splitting the receiver positions into two disjoint sets. The reference and target receiver positions are sampled from the same receiver position set. This means that, during both the training and testing stages, the model will not be trained/tested to predict a warping field that warps audio from a receiver position in the training set to a position in the test set, or vice versa. 

In the synthetic data generation, we arrange the receivers in an $80 \times 40$ grid, with adjacent receivers spaced 0.05 meters apart. To create the test set, we select test samples so that no two test samples are adjacent in the grid. This interleaved sampling strategy ensures that each test receiver position is at least 0.05 meters away from any receiver position in the training set.

In the Photo-realistic data generation, we randomly sample 4500 and 8500 receiver positions on the two scenes' floors and select a subset of 500 receiver positions from each set as the test set. Due to the existence of furniture and other obstacles presented in the room scene, we could not employ the grid-sampling strategy used in synthetic data generation, and could only randomly sample train and test receiver positions on the scene floor. 

To train our model on the real dataset, which has a significantly smaller sampling density required by our model, we first simulate the real dataset scene using the Pyroomacoustic simulator~\cite{pyroomacoustic} and generate 10800 samples in the scene. The receiver positions are arranged in a $90 \times 120$ grid, with adjacent receivers spaced 0.05 meters apart. 1000 test samples are selected from the total 10800 receiver positions. The test set sampling strategy is the same as the interleaved sampling strategy used in the synthetic data generation.

We pretrain our model on the simulated samples before fine-tuning on the real dataset. Receiver positions in the real dataset are also arranged in a grid structure, which allows us to use the interleaved sampling strategy to sample 24 test set receiver positions from the total 130 receiver positions.

\subsection{Effect of Sample Size on Model Training performance}

\begin{figure}
    \centering
    \includegraphics[width=1.1\textwidth]{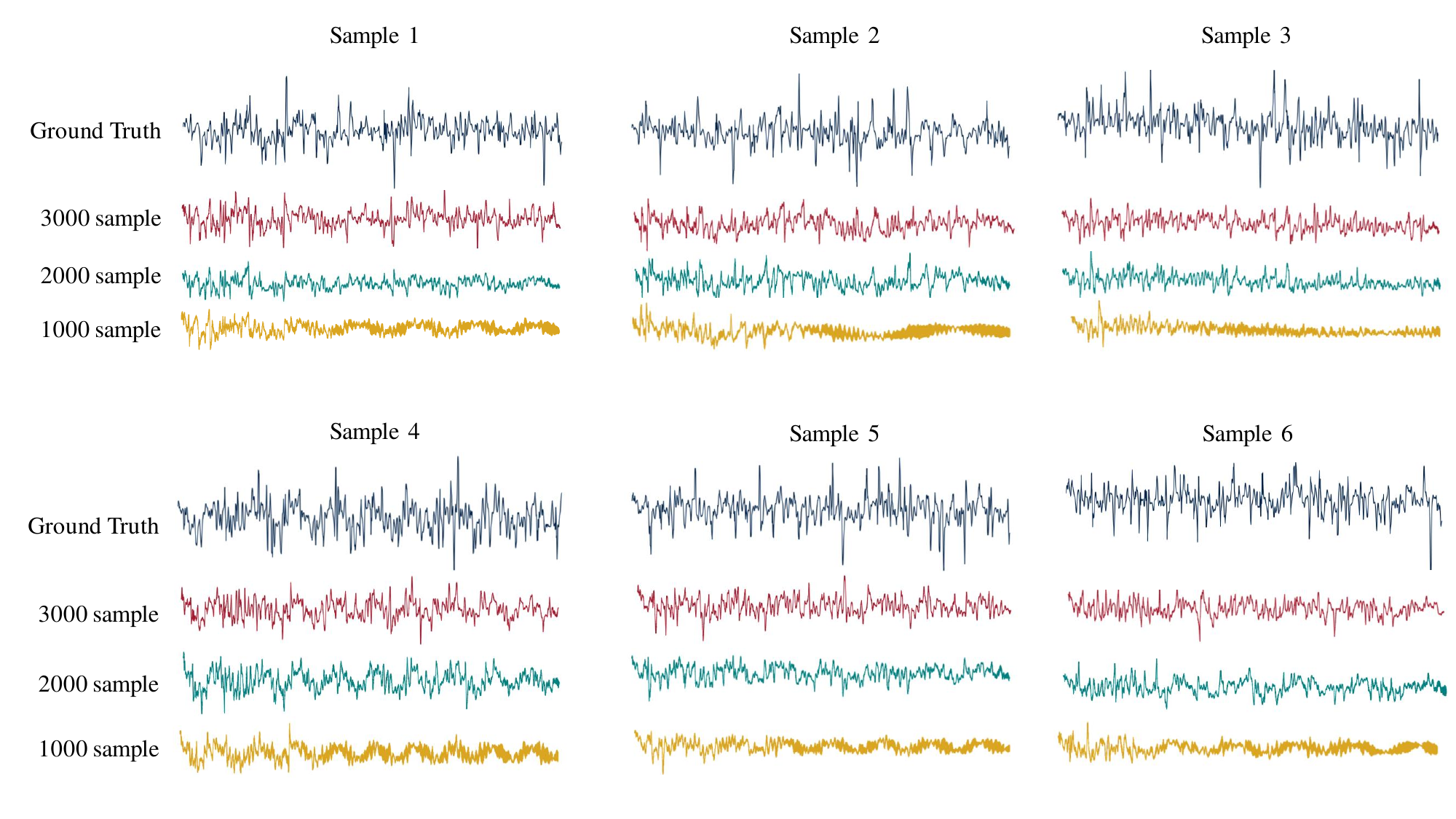}
    \caption{Predictions of models trained with different sample size.}
    \label{fig:sample-size}
\end{figure}

In all three datasets, reference and target audios are densely sampled from the scene. In this section, we show the necessity to sample audio at high density by training our SPEAR model on the same synthetic scene with different sampling densities. We randomly select a subset of 1000 samples and 2000 samples from the whole 3000 synthetic training data samples and show their performance metrics. Tab.~\ref{tab:sample-density} shows the metric of the three models. The model performance drops significantly as the sampling density decreases. In addition, we visualize the predicted warping field of the three models in Fig.\ref{fig:sample-size}. Models trained with smaller sample sizes show significantly worse performance in higher frequency warping field prediction.

\begin{table}[t]
    \centering
    \begin{tabular}{l|cccc}
    \hline
    Sample Size & SDR & MSE & PSNR & SSIM \\
    \hline
    3000 & 1.50 & 0.92 & 15.81 & 0.87 \\
    2000 & 1.06 & 0.96 & 15.27 & 0.87 \\
    1000 & 0.24 & 1.15 & 14.26 & 0.85  \\
    \hline
    \end{tabular}
    \caption{Effect of different sampling density on model performance.}
    \label{tab:sample-density}
\end{table}

\section{\emph{SPEAR} Network Architecture}

\emph{SPEAR} network architecture is shown in Tab.~\ref{spear_network}, and we also provide the code in the suppementary material.

\begin{table}[t]
    \centering
    \begin{tabular}{|c|c|c|}
        \hline
        Layer Name & Filter Num  & Output Size \\
        \hline
        \multicolumn{3}{|c|}{\textbf{Model Input}: 2 position 3d coordinate: [2, 3]}\\
        \hline
        \multicolumn{3}{|c|}{\textbf{Grid Feature}: concatenated 2 position feature: [1, 384]}\\
        \hline
        \multicolumn{3}{|c|}{\textbf{Transformer Encoder Input}: Initial Token Representation: [43, 384]}\\
        \hline
        Transformer Layer 1 & head num = $8$, hidden dim = $384$  & [43, 384]\\
        ... & ... & ... \\
        Transformer Layer 12 & head num = $8$, hidden dim = $384$  & [43, 384]\\
        \hline
        \multicolumn{3}{|c|}{\textbf{Prediction Head}} \\
        \hline
        Real part FC & FC, output\_feat = 384 & [43, 384] \\
        Imaginary part FC & FC, output\_feat = 384 & [43, 384] \\
        \hline
        Flattern & Flattern real/imaginary token sequence. & [16512] \\
         & Construct complex sequence. & \\
        \hline
        Prune & Cut the sequence to 16384 length & [16384] \\
        \hline 
        Mirroring & Generate the full warping field by & [32768] \\
         & concatenating the predicted sequence & \\
         & with its mirrored conjugate sequence & \\
        \hline
    \end{tabular}
        \caption{\emph{SPEAR} Network Architecture Detail.}
    \label{spear_network}
\end{table}

\end{document}